\documentclass[12pt]{article}

\usepackage{amsmath,amssymb,amsfonts,amsbsy}
\usepackage{graphicx}
\usepackage{wrapfig}

\begin{document}
\addcontentsline{toc}{subsection}{{About spin particle solution in Born-Infeld nonlinear electrodynamics}\\
{\it A.A. Chernitskii}}

\setcounter{section}{0}
\setcounter{subsection}{0}
\setcounter{equation}{0}
\setcounter{figure}{0}
\setcounter{footnote}{0}
\setcounter{table}{0}

{

\def\bfgr#1{\boldsymbol{#1}}
\def\cP{{\mathcal{P}}}
\def\cH{{\mathcal{H}}}
\def\br{\bar{r}}
\def\lb#1{\label{#1}}
\def\cbP{{\bfgr{\mathcal{P}}}}
\def\p{\partial}
\def\pp{{\overline\partial}}
\def\dfrac#1#2{{\displaystyle\frac{#1}{#2}}}
\def\be{\begin{equation}}
\def\ee{\end{equation}}
\def\ba#1{\begin{array}{#1}}
\def\ea{\end{array}}
\def\bea{\begin{eqnarray}}
\def\eea{\end{eqnarray}}
\def\df{{\mathrm{d}}}
\def\I{{\mathcal{I}}}
\def\J{{\mathcal{J}}}
\def\div{{\mathrm{div}}}
\def\curl{{\mathrm{curl}}}
\def\Div{{\mathrm{Div}}}
\def\Curl{{\mathrm{Curl}}}
\def\E{{\mathbf{E}}}
\def\B{{\mathbf{B}}}
\def\D{{\mathbf{D}}}
\def\H{{\mathbf{H}}}
\def\L{{\mathcal{L}}}
\def\V{{\mathbf{V}}}

\def\combss#1#2#3{{\mathsurround 0pt\hbox to 0pt {\hspace*{#3}\raisebox{#2}{${\scriptscriptstyle #1}$}\hss}}}
\def\Ce{\combss{e}{0.45ex}{0.2em}C}
\def\Cm{\combss{m}{0.45ex}{0.1em}C}

\def\metrsp{\mathsf{g}}

\begin{center}
\textbf{ABOUT SPIN PARTICLE SOLUTION\\ IN BORN-INFELD\\ NONLINEAR ELECTRODYNAMICS}

\vspace{5mm}

\underline{A.A.~Chernitskii}$^{\,1,2}$

\vspace{5mm}

\begin{small}
  (1) \emph{A. Friedmann Laboratory for Theoretical Physics, St.-Petersburg} \\
  (2) \emph{State University of Engineering and Economics,\\ Marata str. 27, St.-Petersburg, Russia, 191002} \\
  \emph{E-mail: AAChernitskii@mail.ru, AAChernitskii@engec.ru}
\end{small}
\end{center}

\vspace{0.0mm} 

\begin{abstract}
  The axisymmetric static solution of Born-Infeld nonlinear electrodynamics with ring singularity is investigated.
This solution is considered as a static part of massive charged particle with spin and magnetic moment.
The method for obtaining the appropriate approximate solution is proposed. An approximate solution is found.
The values of spin, mass, and magnetic moment is obtained for this approximate solution.
\end{abstract}

\vspace{7.2mm}

The purpose of the present work is construction of the field model for massive charged elementary particle with spin and magnetic moment as
a soliton solution of nonlinear electrodynamics.

In this approach the mass and the spin of the particle have fully field nature and must be calculated with integration of the appropriate densities
over tree-dimensional space.

The charge and the magnetic moment characterize the singularities and the behaviour of the
fields at space infinity.

This theme was discussed in my articles
(see \cite{Chernitskii1999,Chernitskii2004a,Chernitskii2006a,Chernitskii2007a}).

Here we considered the appropriate soliton solution with ring singularity in
Born-Infeld nonlinear electrodynamics.

For inertial reference frames and in the region outside of field
singularities, the equations are
\begin{equation}
\begin{array}{rclrcl}
\mathrm{div} \mathbf{B}  &=& 0\;,& \mathrm{div} \mathbf{D}  &=& 0\;,\\
\dfrac{\partial\mathbf{B}}{\partial t}  {}+{}  \mathrm{curl}\mathbf{E}  &=& 0\;,& \dfrac{\partial\mathbf{D}}{\p t}  {}-{}  \mathrm{curl}\mathbf{H}  &=& 0\;,
\end{array}
\label{Eq:Maxwell}
\end{equation}
where
\begin{align}
\label{BImrel}
& \mathbf{D} = \dfrac{1}{\L}\,(\E  {}+{}  \chi^2\,\J \B)\, ,\quad \mathbf{H} = \dfrac{1}{\L}\,(\B  {}-{}  \chi^2\,\J \E)
\\
\nonumber
&\L {}\equiv{}   \sqrt{|\,1 {}-{}  \chi^2\,\I  {}-{}  \chi^4\,\J^2\,|}\,,\quad \I = \E\cdot\E  {}-{}  \B\cdot\B\,,\quad \J = \E\cdot\B\, .
\end{align}

Here $\E$ and $\mathbf{H}$ are electrical and magnetic field strengths,
$\mathbf{D}$ and $\B$ are electrical and magnetic field inductions.

Mass and spin is defined as three dimensional space integral from the appropriate densities:
\begin{equation}
m=\int\limits_{V}\mathcal{E}\,dv\;,\qquad s = \biggl|\int\limits_{V}\mathbf{r}\times\bfgr{\mathcal{P}}\,dv\biggr|\;,
\end{equation}
where $\bfgr{\mathcal{P}}=\dfrac{1}{4\pi}\mathbf{D}\times\mathbf{B}$ is the Poynting vector.

For Born-Infeld electrodynamics we have the following energy density:
\begin{equation}
\mathcal{E}=\dfrac{1}{4\pi\,\chi^2}\left(\sqrt{1+\chi^2\left(\mathbf{D}^2+\mathbf{B}^2\right)+(4\pi)^2\,\chi^4\,\bfgr{\mathcal{P}}^2}-1\right)\;.
\end{equation}

The behaviour of electrical and magnetic fields for particle solution at infinity is characterized by electrical charge and magnetic moment.
For more details see my paper \cite{Chernitskii2006a}
In spherical coordinates the field components have the following form for $r\!\to\! \infty$:
\begin{align}
\label{EDatinf}
\left\{
D_r,\, D_\vartheta,\, D_\varphi
\right\}
\sim
\left\{
E_r,\, E_\vartheta,\, E_\varphi
\right\}
&\sim
\left\{
\dfrac{e}{r^2},\, 0,\, 0
\right\}
,\\
\label{BHatinf}
\left\{
H_r,\; H_\vartheta,\; H_\varphi
\right\}
\sim
\left\{
B_r,\; B_\vartheta,\; B_\varphi
\right\}
&\sim
\left\{
\dfrac{2\mu\;\cos\vartheta}{r^3},\; \dfrac{\mu\,\sin\vartheta}{r^3},\;  0
\right\}.
\end{align}

The electrical charge and the magnetic moment characterize also the behaviour of fields near singularities.

\begin{wrapfigure}[12]{R}{60mm}
\begin{picture}(100,100)
{\unitlength 0.5mm
\put(0,0){
\begin{picture}(0,0)
  \centering 
  \vspace*{-8mm} 
  \put(0,0){\includegraphics[width=60mm]{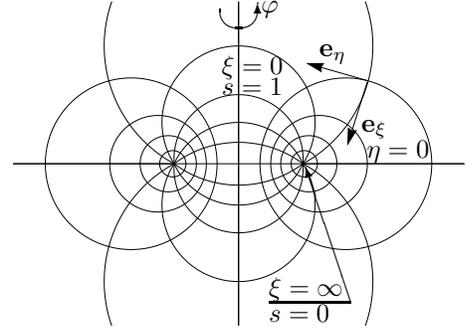}}
\put(60,84){\oval(10,10)[b]}
\put(65,85.5){\vector(0,1){0.1}}
\put(66,84){\makebox(0,0)[lc]{\footnotesize $\varphi$}}
\put(96,52){\makebox(0,0)[cc]{\footnotesize $\mathbf{e}_\xi$}}
\put(85,72){\makebox(0,0)[cc]{\footnotesize $\mathbf{e}_\eta$}}
\put(68,9.5){\makebox(0,0)[lc]{\footnotesize $\xi=\infty$}}
\put(68,3){\makebox(0,0)[lc]{\footnotesize $s=0$}}
\put(68,6){\line(1,0){21.8}}
\put(89.8,6){\vector(-1,3){12}}
\put(55.5,68.5){\makebox(0,0)[lc]{\footnotesize $\xi=0$}}
\put(55.5,63.5){\makebox(0,0)[lc]{\footnotesize $s=1$}}
\put(94,46){\makebox(0,0)[lc]{\footnotesize $\eta=0$}}
\end{picture}
}
}
\end{picture}
  \caption{\footnotesize
The section $(\varphi = 0 \cup \varphi = \pi)$ of the toroidal coordinate system $\{\xi,\eta,\varphi\}$ ($\{s,\eta,\varphi\}$).}
  \label{Chernitskii_fig3}
\end{wrapfigure}

Let us consider the toroid symmetry field configuration.
We use the toroidal coordinate system $\{\xi\in [0,\infty],\eta\in (-\pi,\pi],\varphi\in (-\pi,\pi]\}$ which is
obtained with rotation of bipolar coordinate system around the standing axis.
It is convenient to use the new variable $s\equiv \dfrac{1}{\cosh\xi}\in [0,1]$ instead of the variable $\xi$.

Let us consider the appropriate solution for linear electrodynamics ($\mathbf{D}=\E$, $\mathbf{H}=\B$).

The components of electromagnetic potential for the linear case have the following form:
\begin{align}\label{LinearA0}
    A_0=-\dfrac{e}{\rho _0\,\sqrt{2\,s}}\,\sqrt{1-s\,\cos\eta}\;P_{-\frac{1}{2}}^{0}(1/s)
    \quad,\\
    A_\varphi=-\dfrac{\mu\,2\sqrt{2}}{\rho_0^2\,\sqrt{s}}\,\sqrt{1-s\,\cos\eta}\;i\,P_{-\frac{1}{2}}^{1}(1/s)
    \quad.
    \label{LinearAphi}
\end{align}
where $P_{n}^{m}(x)$ is associated Legendre function.

The electrical and magnetic fields obtained from the potential $\{A_0,0,0,A_\varphi\}$ (\ref{LinearA0}), (\ref{LinearAphi}) has the right behaviour at infinity ($r\!\to\! \infty$ in spherical coordinates) (\ref{EDatinf}).
The behaviour of the field near the singular ring ($s\!\to\! 0$) is the following:
\begin{align}
\label{fEnearRing}
\left\{
E_\xi,\; E_\eta,\; E_\varphi
\right\}
=
\left\{
D_\xi,\; D_\eta,\; D_\varphi
\right\}
&=
\left\{
-\dfrac{e}{\pi\,\rho_0^2\,s},\; 0,\; 0
\right\}+\mathit{o}(s^{-1})
\quad,
\\
\label{fBnearRing}
\left\{
B_\xi,\; B_\eta,\; B_\varphi
\right\}
=
\left\{
H_\xi,\; H_\eta,\; H_\varphi
\right\}
&=
\left\{
0,\; -\dfrac{2\,\mu}{\pi\,\rho_0^3\,s},\; 0
\right\}+\mathit{o}(s^{-1})
\quad.
\end{align}

Let us search  the appropriate solution for Born-Infeld electrodynamics in the following form:
\begin{align}
\label{solBIia1}
A_0 &=-\dfrac{e}{\rho _0\,\sqrt{2}}\,\sqrt{1-s\,\cos\eta}\;f_1(s,\eta)\; ,
\\
A_\varphi &=\dfrac{\mu\,\sqrt{2}}{4\,\rho_0^2}\,\sqrt{1-s^2}\,\sqrt{1-s\,\cos\eta}\;f_2(s,\eta)
\; ,
\label{solBIia2}
\end{align}
where the functions $f_i(s,\eta)$ must be found.

We take the condition of finiteness for the potentials $(A_0,A_\varphi)$ and fields $(\E,\B)$ as physically defensible.

We can seek the functions $f_i(s,\eta)$ in the form of formal double series,
that is the power series in $s$ and the Fourier series in $\eta$. But here we have the problem such that the Lagrangian $\L$
for an appropriate approximate solution
can become
nonanalytic function for sufficiently large values of variable $s$. To avoid the non-analyticity we must take the condition
$(1 {}-{}  \chi^2\,\I  {}-{}  \chi^4\,\J^2)\geq 0$. This condition can be satisfied with the help of exponential smoothing factors.
Thus let the appropriate approximate solution be have the following form:
\begin{align}
\label{potEBsol}
    f_i(s,\eta)\approx f_i^{NK}&=
    c_{i0}\,\mathrm{e}^{-q_{i0}\, s^{N+1}}+\sum\limits_{n=1}^{N}\sum\limits_{k=0}^{K}c_{ink}
    \,\mathrm{e}^{-q_{ijk}\, s^{N+1}} s^n\,\cos k\eta
    \; ,
\end{align}
where the integers $N$ and $M$ characterize the order of the approximation, $q_{i0}\geq 0$, $q_{ink}\geq 0$.

Then we substitute the field functions $\{\E,\B\}$ obtained from the potentials (\ref{solBIia1}), (\ref{solBIia2}) with the functions
(\ref{potEBsol}) to equations
\begin{equation}\label{divDeq0curlHeq0}
\div \D=0\,,\quad \curl \mathbf{H}=0
\end{equation}
with Born-Infeld relations $\mathbf{D} (\mathbf{E},\mathbf{B})$ and $\mathbf{H} (\mathbf{E},\mathbf{B})$
(\ref{BImrel}).

We have the fields near $s=0$ in the following form:
\begin{align}
\label{fEBnearRingBI}
\left\{
E_\xi,\; E_\eta,\; B_\xi,\; B_\eta
\right\}
&=
\left\{
-\dfrac{e\,\sqrt{\rho_0^4+\beta^2}}{\rho_0^2},\; 0,\; 0,\; -\dfrac{e\,\beta}{\rho_0^2}
\right\}+\mathcal{O}(s)
\quad,
\\
\label{fDHnearRingBI}
\left\{
D_\xi,\; D_\eta,\; H_\xi,\; H_\eta
\right\}
&=
\left\{
-\dfrac{e}{\pi\,\rho_0^2\,s},\; 0,\; 0,\; -\dfrac{e\,\beta}{\pi\,\rho_0^2\,s\,\sqrt{\rho_0^4+\beta^2}}
\right\}+\mathcal{O}(1)
\quad,
\end{align}
where $\beta = \mu\left/\dfrac{e\,\rho_0}{2}\right.$.

As we can see the fields $\{\E,\B\}$ near $s=0$ is finite and the fields $\{\D,\H\}$ have the behaviour of the same type that in the linear
case (\ref{fEnearRing}), (\ref{fBnearRing}).

The extraction of the coefficients for $s^n$ and $k$-th Fourier harmonic in equations (\ref{divDeq0curlHeq0})
allows to have the linear systems for obtaining the coefficients $c_{ink}$.

The coefficients $c_{i0}$ must be obtained from the conditions
\begin{align}\label{fieq1}
    f_i(1,0)=1
\end{align}
These relation satisfy the conditions (\ref{EDatinf}) and (\ref{BHatinf}) at space infinity ($r\!\to\!\infty$).

The approximate solution of the form (\ref{potEBsol}) was found for $N=3$ and $K=3$.

The obtained approximate solution in appropriate
units ($e=1$, $r_0=\sqrt{|e\,\chi|}=1$) has the following free parameters: radius of the ring $\rho_0$,
magnetic moment $\mu$, smoothing coefficients $q_{i0}$, $q_{ijk}$.

But the solution of Born-Infeld electrodynamics under investigation should not have continuous free parameters except for the electrical charge.
Thus we must have a set of free parameters appropriate for the best approximation to solution.

The appropriate values of the free parameters was found by the direct numerical minimization of the action functional
$\int(\L-1) dv$ with the condition $(1 {}-{}  \chi^2\,\I  {}-{}  \chi^4\,\J^2)\geq 0$.

The action with the approximate solution under consideration reach a local minimum for the following values:
\begin{align}
\label{rhobe}
\rho_0\approx 0.95 \,r_0\quad,\qquad\mu\approx 0.80\,\dfrac{e\,\rho_0}{2}\quad.
\end{align}
We have the following mass and spin for the obtained field configuration:
\begin{align}
\label{ms}
m\approx \dfrac{10}{\chi^2}\quad,\qquad s\approx 0.1\,\dfrac{e^2}{2\alpha}\quad,
\end{align}
where $\alpha$ is the fine structure constant and $e^2/(2\alpha) = \hbar/2$.

It should be noted that the value $\mu\sim e\,\rho_0/2$ can be considered as desirable for this model (see my artice \cite{Chernitskii2008a}).

Thus the obtained result is encouraging. But of course it must be considered as preliminary.
}

{
\expandafter\ifx\csname natexlab\endcsname\relax\def\natexlab#1{#1}\fi
\providecommand{\enquote}[1]{``#1''}
\expandafter\ifx\csname url\endcsname\relax
  \def\url#1{\texttt{#1}}\fi
\expandafter\ifx\csname urlprefix\endcsname\relax\def\urlprefix{URL }\fi
\providecommand{\eprint}[2][]{\url{#2}}

}

\end{document}